\documentclass{webofc}
\usepackage[varg]{txfonts}   
\usepackage{hyperref}
\usepackage{url}
\usepackage{graphicx}
\usepackage[figuresright]{rotating}
\usepackage{makecell}
\usepackage{bm,amsmath,amssymb}
\usepackage[mathscr]{eucal}
\usepackage{multirow}
\usepackage{xcolor}
\usepackage{mathrsfs}
\usepackage{mathtools}
\usepackage{slashed}
\usepackage{graphics}
\usepackage{graphicx}
\usepackage{subfigure}
\usepackage{dsfont}
\usepackage{longtable}
\usepackage{bbm} 
\usepackage{float}
\usepackage{tcolorbox}
\usepackage{lipsum}
\usepackage{cancel}
\usepackage{enumerate}
\usepackage{diagbox}

\hypersetup{colorlinks=true,citecolor=blue,urlcolor=blue,linkcolor=blue}
\newcommand{\NN}{\mathrm{NN}}
\newcommand{\rmd}{{\rm{d}}}

\begin{document}
\title{Open heavy-flavor transport and hadronization \\ 
in heavy-ion collisions}
%
%

\author{
\firstname{Yu} \lastname{Fu}\inst{1}\fnsep\thanks{\email{yu.fu@duke.edu}} 
\and
\firstname{Tharun} \lastname{Krishna}\inst{2}
\and\firstname{Weiyao} \lastname{Ke}\inst{3}
\and
\firstname{Steffen} \lastname{A. Bass}\inst{1}
\and
\firstname{Ralf} \lastname{Rapp}\inst{2}
}

\institute{Department of Physics, Duke University, Durham, NC 27708-0305, U.S.A 
\and
Department of Physics and Astronomy and Cyclotron Institute, Texas A\&M University,\\
College Station, TX 77843-3366, U.S.A
\and
Key Laboratory of Quark and Lepton Physics (MOE) \& Institute of Particle Physics,\\
Central China Normal University, Wuhan 430079, China
          }

\abstract{
We develop a comprehensive model for heavy-quark evolution in a realistic QGP, from their production in the initial collision to hadronic freeze-out. Heavy-quark transport is described by a Langevin approach including medium-induced radiation, coupled to a 2+1D viscous hydrodynamic bulk evolution. Transport coefficients are obtained from non-perturbative $T$-matrix calculations with resonant correlations near the transition temperature. Hadronization is implemented via two fragmentation+recombination schemes: an improved sudden coalescence model and a resonance recombination model. We present results for key open heavy-flavor observables, i.e., the nuclear modification factor and elliptic flow, and compare to LHC Pb-Pb data at $\sqrt{s_{\NN}}$=5.02\,TeV.}
\maketitle
\section{Introduction}
\label{sec:intro}
Heavy quarks (HQs) are pristine probes of the quark-gluon plasma (QGP) created in ultrarelativistic heavy-ion collisions (URHICs). Owing to their large mass, they are predominantly produced in the earliest stages of the collision, with rates calculable in perturbative QCD. Their comparatively long thermal relaxation time enables them to preserve much of their interaction history with the medium, including transport in the QGP, hadronization, and subsequent rescattering in hadronic matter. Consequently, HQs carry valuable information about the entire evolution of the QGP.

Over the past two decades, considerable effort has been devoted to modeling HQ transport in the QGP. Despite this, a fully comprehensive framework combining nonperturbative diffusion coefficients, realistic in-medium radiation, energy-momentum conserving hadronization, hadronic-phase diffusion, and viscous hydrodynamic evolution with realistic initial conditions remains elusive. In this work, we report progress toward such a framework by combining: (1) a bulk medium constrained via Bayesian analysis of light-hadron observables, (2) HQ transport with nonperturbative $T$-matrix diffusion coefficients and Boltzmann radiation with improved interference effects, (3) state-of-the-art hadronization with recombination that conserves energy-momentum supplemented by fragmentation, and (4) elastic hadronic rescattering.

\section{Framework}
\label{sec:framework}
In this section, we summarize the main ingredients of the framework mentioned above.

\subsection{Bulk evolution model}
The bulk medium evolution is initialized through the deposition of energy and density in the earliest stage of the collision. This stage is modeled utilizing the initial-condition model T$_\text{R}$ENTo~\cite{Moreland:2014oya}. The system then undergoes free-streaming for a proper time $\tau_{fs}$ to mimic QGP thermalization, serving as the input for the subsequent hydrodynamic evolution. The dynamical expansion of the thermalized QGP is described using the (2+1)-dimensional viscous hydrodynamics model VISHNU~\cite{Song:2007ux}. All parameters have been determined via Bayesian analysis of light-hadron observables.

\subsection{Heavy quark transport in the QGP}
Heavy quarks, as probes of the bulk medium's properties, are embedded in and evolve alongside the QGP. The hydrodynamic simulation supplies local velocity and temperature profiles for the dynamic interactions between the HQs and the medium. The interactions are encapsulated in transport coefficients which are discussed in Sec.~\ref{ssec:Tmatrix}.

The HQ initial conditions are specified by momentum spectra taken from proton-proton collisions sampled from a FONLL calculation and scaled by the number of binary collisions in nucleus–nucleus systems. In coordinate space, the initial HQ positions are distributed according to the binary-collision profile provided by T$_\text{R}$ENTo.

The dynamic evolution of HQs in the QGP is described by the LIDO model~\cite{Ke:2018tsh,Ke:2018jem}, which combines Langevin diffusion with medium-induced radiation, $\partial_t f_Q = \mathcal{D}[f_Q] + \mathcal{C}_{1\to 2}[f_Q]$,
where $\mathcal{D}$ is the Fokker-Planck diffusion operator and $\mathcal{C}_{1\to2}$ accounts for diffusion-induced radiation. 
The diffusion process is realized via the pre-point Langevin equation,  
$
\bm{p}(t+\Delta t) = \bm{p}(t) - \eta_D \bm{p}\,\Delta t + \bm{\xi}(t)\Delta t \ ,
$
where the HQ-medium interactions are encoded in the drag coefficient, $\eta_D$, and momentum diffusion coefficient, $\kappa$, related through the Einstein relation via $\kappa=2TE\eta_D$. 
The random force satisfies $\langle \xi_i(t)\xi_j(t+\Delta t) \rangle = \kappa \delta_{ij}/\Delta t$. 
The spatial diffusion coefficient and jet transport coefficient are given by $D_s=2T^2/\kappa$ and $\hat{q}_S=2\kappa$, respectively. The diffusion-induced radiation component is calculated via elastic collision rates from the same underlying transport coefficient. For 1$\to$2 processes, the rate is given by $  R_{1\to2}= \hat{q}_S \int\rmd x \int\rmd k_{\perp}^2 \frac{\alpha_s(k_{\perp}^2) P(x)}{2\pi(k_{\perp}^2 +m_{\infty}^2)}$,
where $P(x)$ is the vacuum QCD splitting function and $m_{\infty}^2=m_D^2/2$ is the asymptotic thermal gluon mass. We adopt a running coupling in the form $\alpha_s(Q^2) = \frac{4\pi}{9\ln[\max\{\,Q^2,(\mu \pi T)^2\,\}/\Lambda^2] }$, with $\Lambda=0.2~\mathrm{GeV}$ and a dimensionless parameter, $\mu$, controlling the medium scale.

\subsection{Heavy quark medium interaction}
\label{ssec:Tmatrix}
\begin{figure}[H]
\centering
\sidecaption
\includegraphics[width=3.5cm,clip]{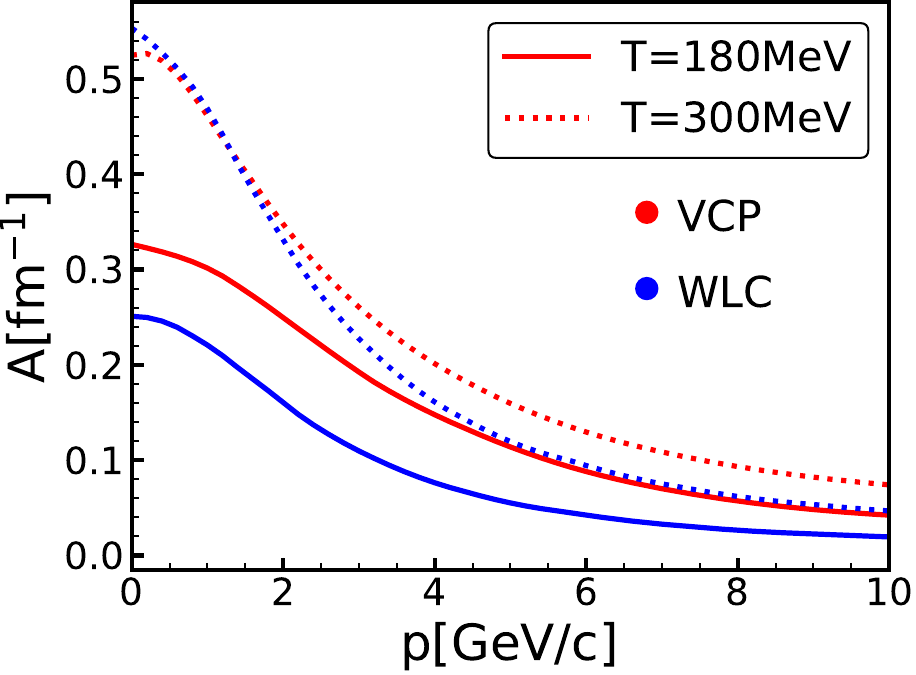}
\includegraphics[width=3.5cm,clip]{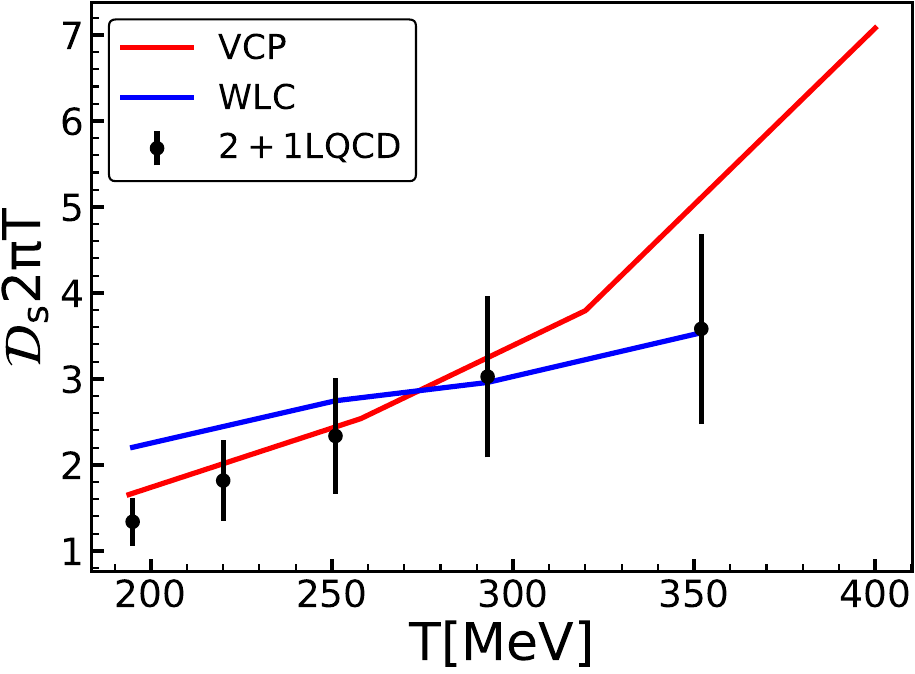}
\caption{Left: Friction coefficient for charm quarks from $T$-matrix calculations in the VCP (red) and WLC (blue) scenarios at $T$=180\,MeV (solid) and 300\,MeV (dotted). Right: Spatial diffusion coefficient for charm quarks in comparison to recent lattice data~\cite{Altenkort:2023oms,Altenkort:2023eav}.}
\label{fig:ApandDs}     
\end{figure}
The transport coefficients encode the interactions of HQs with the QGP. In the $T$-matrix framework, these interactions are treated nonperturbatively by resumming ladder diagrams of an in-medium potential, solving one- and two-body correlations selfconsistently with full off-shell dynamics, and treating scattering and bound states on equal footing, and linking HQ transport to the QGP equation of state~\cite{Liu:2016ysz}. After partial-wave expansion, the HQ interaction with thermal partons can be reduced to a 1D integral equation, $T = V + V\otimes G^0\otimes T$, with $G^0$ the in-medium two-parton propagator and $V$ the potential, modeled after a Cornell-type form including Coulomb and string-breaking terms. We focus on two recent implementations of lattice-QCD (lQCD) constraints: (a) \text{Vector confinement potential (VCP)}: constrained by HQ free energies with $1/M$ spin corrections from vacuum spectra, requiring a vector component in the string force which enhances the interaction strength at high momentum~\cite{Tang:2023lcn}; (b) \text{Wilson line correlators (WLC)}: as recently computed in QCD results, which are sensitive to the imaginary part of the potential, yielding a weaker screening at high temperature and a reduced vector admixture ($\sim 20\%$) while still reproducing vacuum spectroscopy~\cite{Tang:2023tkm}.

From the heavy-light $T$-matrices (including all available spin and color channels), the HQ friction coefficient, $A(p,T)$, can be calculated by a convolution over thermal distributions with an angular weighting. In Fig.~\ref{fig:ApandDs} left, we show $A(p)$ for charm quarks at $T=300$~MeV and $T=180$~MeV: it decreases with momentum, with the VCP generally yielding larger values than WLC, indicating stronger coupling to the QGP. The corresponding spatial diffusion coefficients, $D_s(T) = T/[M A(T)]_{p=0}$, are shown in the right panel. Compared to recent lQCD results~\cite{Altenkort:2023oms,Altenkort:2023eav}, the VCP scenario agrees better at low temperature, while the WLC better reproduces the mild temperature dependence suggested by the lattice.

\subsection{Heavy quark hadronization}
\label{sec:hadronization}
As the QGP cools toward the transition temperature, HQs hadronize via fragmentation and recombination. We discuss two sets of hadronization models:

(1) improved Sudden Coalescence Model (iSCM) + Peterson Fragmentation Function (FF): In the SCM, HQs hadronize by recombining with thermal light quarks according to Wigner functions derived from bound-state wave functions, which encode the probability for partons to form hadrons. The improved version~\cite{Cao:2019iqs} extends the hadronic spectrum to include both $S$- and $P$-wave states with full spin and spin–orbit couplings, covering most states listed by the Particle Data Group
(PDG), and implements energy–momentum conservation through the decay of virtual states into a pion and an on-shell HF hadron. HQs that do not recombine hadronize via Peterson fragmentation.

(2) Resonance Recombination Model (RRM) + Heavy Quark Effective Theory (HQET) FF: RRM models HF meson formation via resonant $q+\bar{q}\to M$ scattering using a relativistic Breit–Wigner cross section, derived from evaluating recombination and dissociation rates near equilibrium. The latest version~\cite{He:2019vgs} treats off-equilibrium HQs by computing a self-consistent recombination probability in the local fluid frame, partitioning each HQ between recombination and a HQET-based fragmentation. The model includes all PDG-listed charmed hadrons and additional baryons predicted by the Relativistic Quark Model.

\setlength{\parskip}{0pt}
\section{Results}
In this section, we present and discuss our results for basic HF observables in Pb–Pb collisions at a center-of-mass energy of $\sqrt{s_{\mathrm{NN}}} = 5.02$ TeV for the 30-50$\%$ centrality class, obtained within the framework we have developed here, focusing on $D$ mesons. It should be noted that the present results do not account for hadronic rescattering effects.

\begin{figure}[H]
\centering
\includegraphics[width=6.45cm]{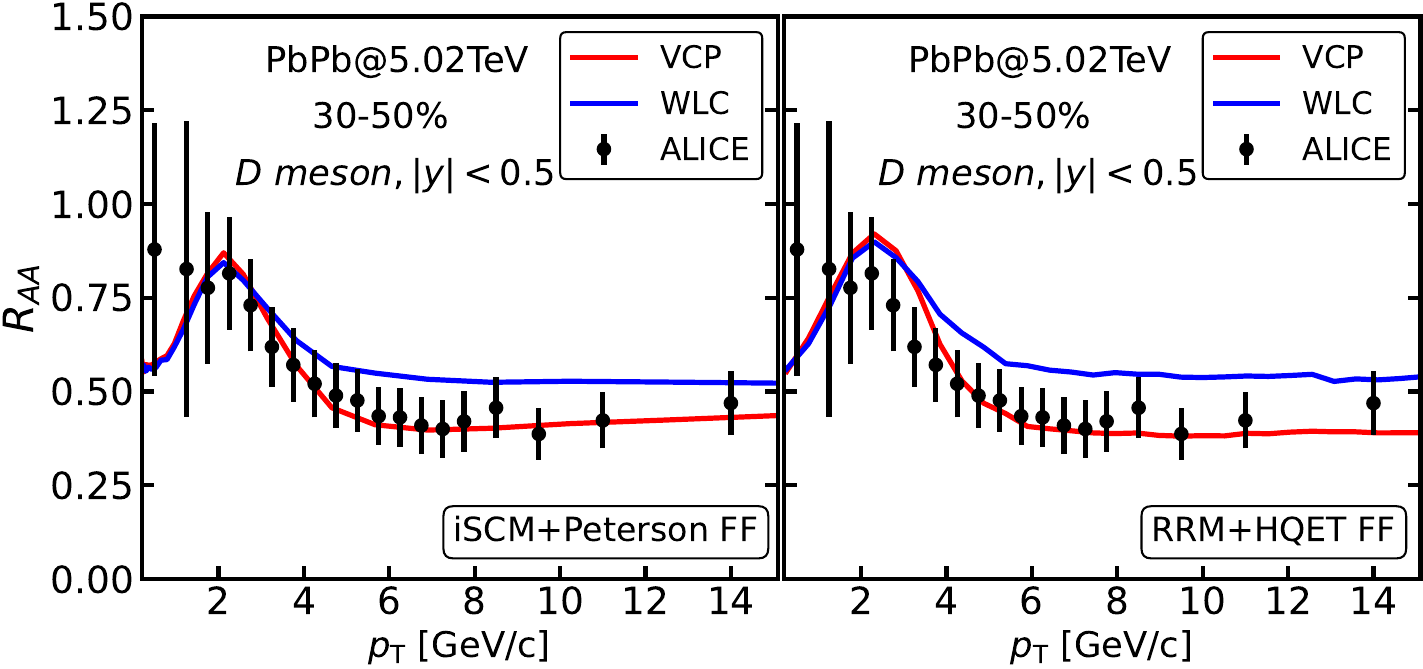}
\includegraphics[width=6.45cm]{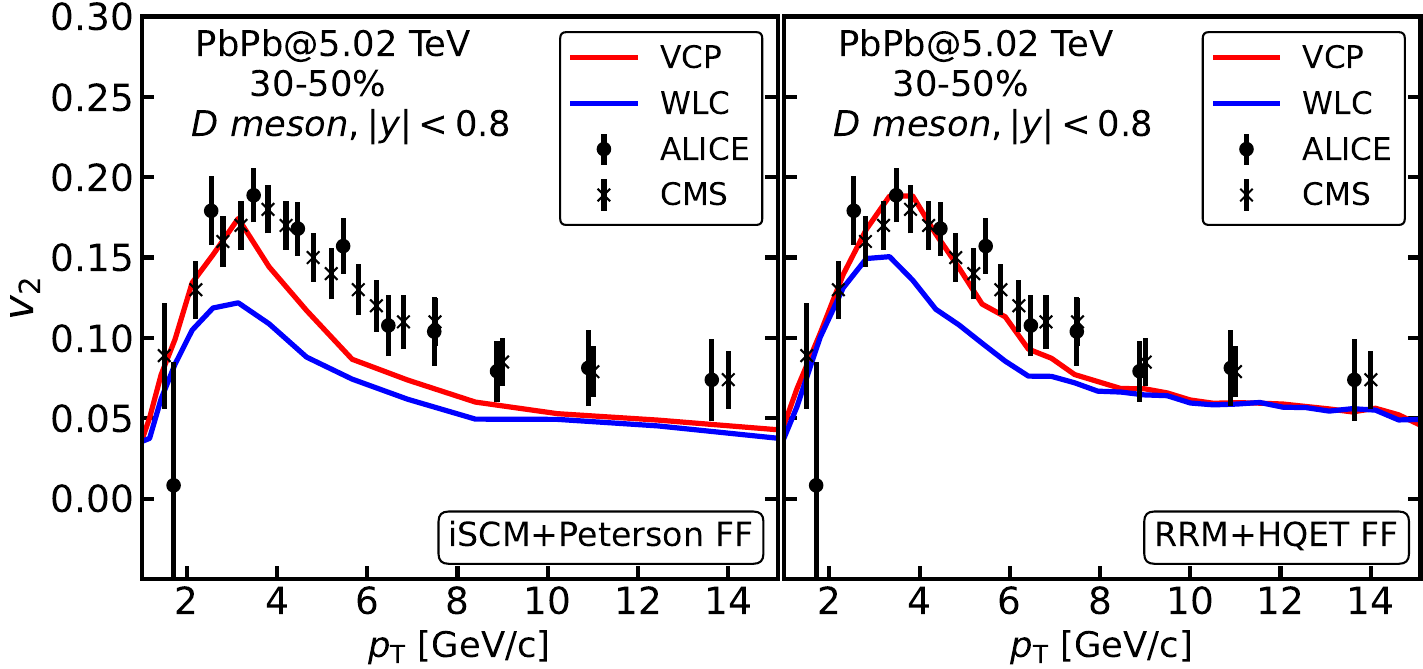}
\caption{Nuclear modification factor (right 2 panels) and elliptic flow (left 2 panels) of $D$-mesons at central rapidity in 30-50\% central Pb-Pb($\sqrt{s_{\NN}}=5.02$~TeV) collisions, compared to ALICE~\cite{ALICE:2021rxa,ALICE:2020iug} and CMS~\cite{CMS:2020bnz} data. Red and blue curves correspond to calculations with VCP and WLC potentials, respectively. Results from different hadronization models are compared in the subpanels.}
\label{fig:RAAandv2}     
\end{figure}

In the left panels of Fig.\ref{fig:RAAandv2} we show the results for the $R_{AA}$ of $D$ mesons compared to corresponding ALICE data~\cite{ALICE:2021rxa}. We observe significant sensitivity to the microscopic interactions from the different in-medium potentials. For both hadronization models, the data favor VCP over WLC. The $D$-meson $v_2$ results (right panels in Fig.~\ref{fig:RAAandv2}), compared to ALICE~\cite{ALICE:2020iug} and CMS~\cite{CMS:2020bnz} data, exhibit a larger sensitivity to the in-medium QCD force at lower $p_{\rm T}$: the VCP scenario captures the experimental trends better than the WLC which leads to systematically smaller flow values.  

\section{Summary}
We have introduced a comprehensive state-of-the-art framework for HQ evolution in QGP. Our results demonstrate sensitivity to the microscopic HQ medium interactions and support the hadronization models that were used. The preliminary data-theory comparison is promising. The incorporation of hadronic transport is ongoing. In future work, we will also extend the present framework to event-by-event simulations and investigate the bottom sector. 

\!\!\!\!\!\!\!\!\!\textbf{Acknowledgment}: This work was supported by the U.S. Department of Energy through the Topical Collaboration in Nuclear Theory on {\it Heavy-Flavor Theory (HEFTY) for QCD Matter} under award no. DE-SC0023547 and award no. DE-FG02-05ER41367 (SAB), as well as by the U.S. National Science Foundation under grant no.~PHY-2209335 (RR).


\end{document}